\documentclass[sigconf]{acmart}

\AtBeginDocument{%
  }

\setcopyright{acmlicensed}
\copyrightyear{2018}
\acmYear{2018}
\acmDOI{XXXXXXX.XXXXXXX}

\acmConference[Conference acronym 'XX]{Make sure to enter the correct
  conference title from your rights confirmation emai}{June 03--05,
  2018}{Woodstock, NY}
\acmISBN{978-1-4503-XXXX-X/18/06}

\def\sysname{ERCache}

\begin{document}
\title{\sysname: An Efficient and Reliable Caching Framework for Large-Scale 
User Representations in Meta's Ads System}

\author{Fang Zhou, Yaning Huang, Dong Liang, Dai Li, Zhongke Zhang, Kai Wang,
Xiao Xin, Abdallah Aboelela, 
Zheliang Jiang, Yang Wang, Jeff Song, Wei Zhang, Chen Liang,
Huayu Li, ChongLin Sun, Hang Yang, Lei Qu, Zhan Shu, Mindi Yuan, Emanuele Maccherani,
Taha Hayat, John Guo, Varna Puvvada, and Uladzimir Pashkevich}

\affiliation{%
  \institution{Meta Platforms, Inc.}
  \city{Menlo Park}
  \state{CA}
  \country{USA}
}

\renewcommand{\shortauthors}{Trovato et al.}

\begin{abstract}
The increasing complexity of deep learning models 
used for calculating user representations presents 
significant challenges, 
particularly with limited computational resources 
and strict service-level agreements (SLAs). 
Previous research efforts have focused on 
optimizing model inference but 
have overlooked a critical question: 
is it necessary to perform user model 
inference for every ad request 
in large-scale social networks?

To address this question and these challenges, 
we first analyze user access patterns at Meta and 
find that most user model inferences occur 
within a short timeframe. T
his observation reveals a triangular relationship 
among model complexity, embedding freshness, 
and service SLAs.

Building on this insight, 
we designed, implemented, and evaluated ~\sysname, 
an efficient and robust caching framework 
for large-scale user representations 
in ads recommendation systems 
on social networks. 
~\sysname{} categorizes cache 
into direct and failover types and 
applies customized settings and eviction policies 
for each model, 
effectively balancing model complexity, 
embedding freshness, and service SLAs, 
even considering the staleness introduced by caching.

~\sysname{} has been deployed at Meta for over six months,
supporting more than 30 ranking models 
while efficiently conserving computational resources 
and complying with service SLA requirements.
\end{abstract}



\maketitle

\section{Introduction}
Deep learning techniques have been shown to significantly improve user representation in recommendation systems
\cite{deep_wide, deepFM, dcn, dcnv2, neural_cf}. 
By leveraging neural networks and other deep learning architectures, 
these models can learn complex patterns and relationships between users and items, 
resulting in more accurate recommendations. 
Therefore, there has been a growing trend towards developing 
increasingly complex deep learning models 
to enhance the performance of user representation
~\cite{deep_wide, fm1, fm2, collaborativefilter, dcn, dcnv2, 
deepFM, airbnb, tencent, pinnerformer, ali, pinsage, twhin, 
pinterest_transact, yuan1, yuan2, zhang2024scaling}.

Since user representation is inferred through online serving,
the increasing complexity of models in ads recommendation systems 
has significant challenges: constrained computational resources
and service SLA limitations.

These challenges necessitate the development of more efficient computational strategies 
and robust system architectures to ensure that 
the deployment of complex models does not 
compromise user experience, recommendation performance, and system reliability.

Prior to our work, researchers have focuses more on
how to speed up the model inference requests,
using scalable embedding structures~\cite{kurniawan2023evstore,pan2023recom},
heterogeneous caching embeddings~\cite{xie2022fleche, song2023ugache}, etc.
However, no prior work has attempted to investigate and understand 
whether it is necessary to perform model inference for every ads request
in large-scale social network.
Our investigation into user access patterns reveals that 
76\% of consecutive user tower inferences occur within ten minutes, 
and 52\% occur within one minute. 
This observation highlights the potential benefits of 
using cached user embeddings to reduce the 
number of requests for model inference.
In addition, it reveals a crucial triangular relationship 
between user embedding freshness, model complexity, 
and service SLAs in ads recommendation systems. 
This interplay highlights the need for a balanced approach 
that takes into account the trade-offs 
between these factors to achieve 
optimal system performance and efficiency.

To address these challenges, we propose \sysname, 
an efficient and reliable caching framework specifically 
designed for large-scale user representation within ads recommendation systems.
Our approach is based on the observation that consecutive user tower inferences 
often occur within a short time frame. 
The primary goal of \sysname{} is 
to achieve an optimal balance between user embedding freshness,
model complexity, and service SLAs.


\sysname{} includes two components: direct cache and failover cache.
The direct cache stores generated user tower embeddings to bypass user tower inference requests when cached embeddings are valid. 
Conversely, the failover cache is employed to recover from failed requests.
We carefully design cache requests and 
eviction policies with customized settings for each model. 

\sysname{} has been deployed at Meta for more than half a year and 
supported more than 30 ranking models at Meta successfully, 
significantly unblocking computational resources limitations 
while adhering to strict service SLAs.

In summary, this paper makes the following contributions:
\begin{enumerate}
    \item [1.] It shows the observation of user access pattern of 
    large-scale social network in ads recommendation,
    which highlights the user access pattern in ads recommendation systems, revealing that a significant portion of consecutive user tower inferences occur within a short time frame.
    \item [2.] It reveals a crucial triangular relationship 
    between user embedding freshness, model complexity, 
    and service SLAs in ads recommendation systems. 
    This interplay highlights the need for a balanced approach 
    that takes into account the trade-offs 
    between these factors to achieve 
    optimal system performance and efficiency.
    \item [3.] It proposes \sysname, a novel caching framework 
    specifically designed for large-scale user representations 
    in ads recommender systems. 
    \sysname{} is a comprehensive solution that addresses 
    the challenges of constrained computational resources
    and service SLA limitations.
    \item [4.] We integrate \sysname{} with 
    a diverse range of ranking models, 
    demonstrating its versatility and adaptability 
    in various ads recommendation scenarios. 
    Through extensive experiments, 
    we show the effectiveness of \sysname{} in 
    improving system efficiency and reliability.
\end{enumerate}

\section{Motivation}
This section presents some challenges and opportunities 
that motivates our work.

\subsection{Challenges}
The complexity of models used in ad recommendation systems 
is increasing at a faster rate than the available computational resources, 
creating challenges in terms of 
constrained computational resources
and 
service SLA limitations.

\begin{figure}[tbp]
  \centering
  \includegraphics[width=.45\textwidth]{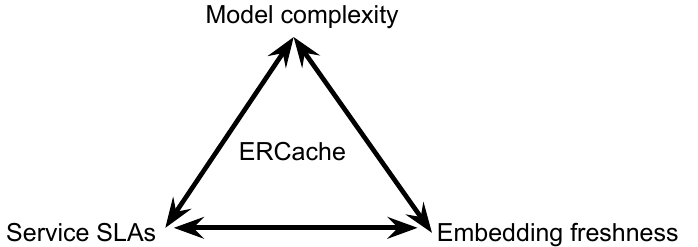}
  \caption{Model Serving Triangle.}
  \label{fig:triangle}
\end{figure}

\textbf{Constrained computational resources:} 
    Increased model complexity raises demand for 
    computational resources (e.g., CPUs, GPUs). 
    However, the availability of these resources is limited in reality, 
    thus not all models' computational needs can be satisfied.
    
\textbf{Service SLA limitations:}
    Incorporating complex models may 
    increase e2e latency and are vulnerable to failures 
    due to computational demands. 
    This could potentially violate service SLAs.

\subsection{Model Serving Triangle}
As shown in Figure~\ref{fig:triangle}, we have observed
a triangular relationship in model serving practice that
it is impossible for a model serving system to
simultaneously provide all three of the following guarantees:
\begin{itemize}
    \item Model complexity: the computation resource required 
    by ML models used in the model serving system.
    \item Embedding freshness: how up-to-date the embeddings 
    (i.e., user tower embeddings) are in the model serving system.
    \item Service SLAs: the requirements of important system metrics,
    like e2e latency, model fallback rate, etc.
\end{itemize}

In other words, if a model serving system is designed to 
handle complex models and provide fresh embeddings, 
it may compromise on service SLAs. 
Similarly, if the system prioritizes embedding freshness 
and meets the requirements of service SLAs, 
it may sacrifice model complexity. 
Alternatively, if the system aims to achieve 
both model complexity and service SLAs, 
it may not be able to maintain embedding freshness.

Since model complexity tends to increase over time, 
and service SLAs remain unchanged in production, 
it is essential to explore opportunities for improvement 
in embedding freshness. This can help maintain a balance 
between the three factors and ensure that the model serving 
system continues to perform optimally.

\subsection{Opportunities}
\begin{figure}[tbp]
  \centering
  \includegraphics[width=.45\textwidth]{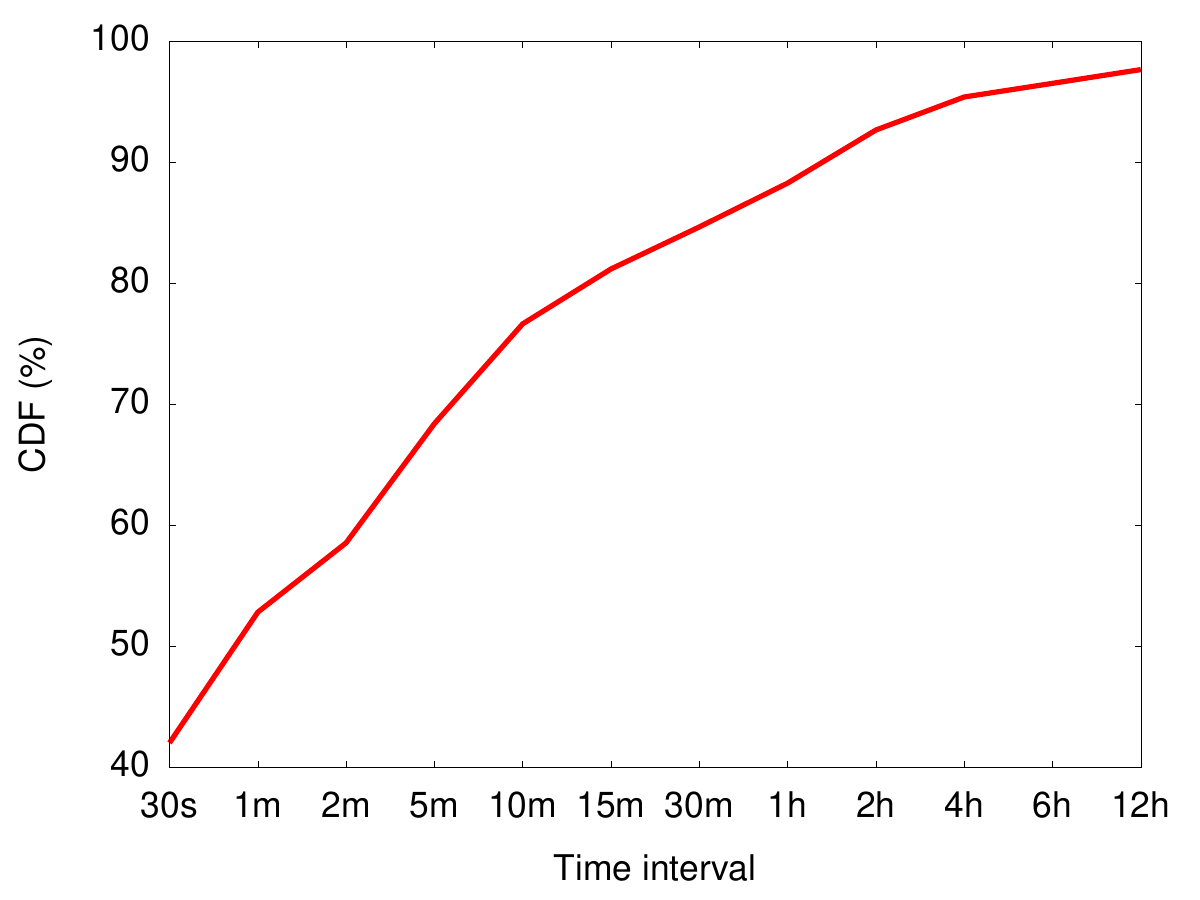}
  \caption{CDF of consecutive inference time interval.}
  \label{fig:cdf-user}
\end{figure}

\begin{figure*}[!h]
  \centering
  \includegraphics[width=.9\textwidth]{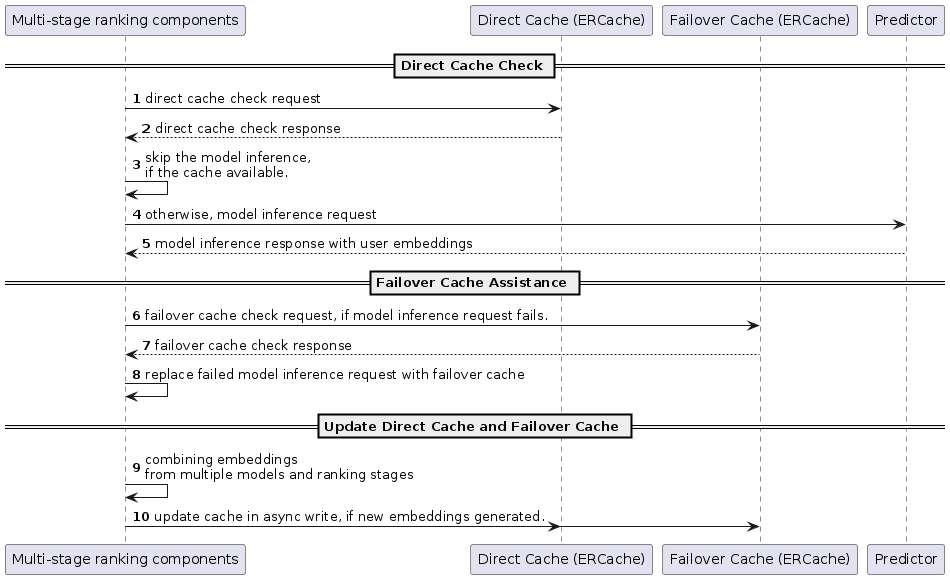}
  \caption{Sequence diagram of \sysname{}.}
  \label{fig:flow}
\end{figure*}

To find the opportunities, 
we review the access pattern for online users interacting
with ads recommendation systems at Meta.
As shown in Figure~\ref{fig:cdf-user},
there is a significant likelihood of multiple user tower inferences 
occurring at a short time.
These findings have motivated us to design a caching system 
to balancing user embedding freshness
and model performance
that takes advantage of user access patterns.

Specifically, 88\% of consecutive user tower inferences 
occur within an hour, 
while 76\% occur within ten minutes and 52\% occur within one minute.

Our observations indicate that 
even short-lived caches, such as one minute, 
can significantly reduce computational resource usage. 
Additionally, mid-lived caches lasting about an hour 
can cover the majority of requests and 
serve as a potential source for failure recovery. 
These findings have motivated us to 
design a caching system 
that balances user embedding freshness, 
model complexity, and service SLAs 
by leveraging user access patterns.

\section{Design and Implementation}
In this section, we introduce the design details of \sysname.

\subsection{Architecture of \sysname.}
\begin{figure}[tbp]
  \centering
  \includegraphics[width=.45\textwidth]{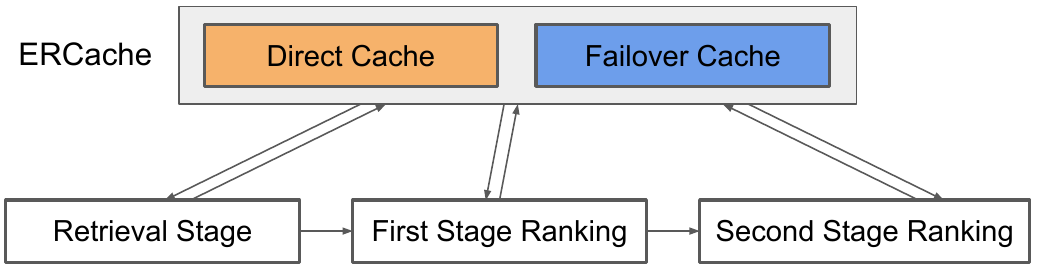}
  \caption{Architecture of the \sysname{} system.}
  \label{fig:scale}
\end{figure}

\sysname{} is a caching system independent from
ads ranking systems, shown in Figure~\ref{fig:scale}.
\sysname{} consists of two components: 
direct cache and failover cache.

The direct cache stores generated user embeddings to 
bypass user tower inference requests when cached embeddings are valid. 
The failover cache applies the cached user embeddings 
to recover from failed requests.

\subsection{\sysname{} functionalities}
\sysname{} offers three functionalities to 
enhance the efficiency and robustness of ads ranking systems:

\begin{itemize}
    \item [1.] Direct Cache Check: System checks if model's direct cache 
    is valid before sending requests to inference; 
    uses cached embedding if valid, otherwise continues normally.
    \item [2.] Failover Cache Assistance: For failed inference requests, 
    system checks failover cache for valid embeddings; 
    replaces failed requests with valid cached embeddings, 
    otherwise reports failure.
    \item [3.] Cache update: Upon receiving the latest embeddings 
    from the model inference requests, 
    we will update the cache in \sysname{} 
    by issuing a write request.
\end{itemize}

The sequence diagram of \sysname{} is shown in Figure~\ref{fig:flow}.

\begin{table*}[htbp]
\centering
\begin{tabular}{ccc}
\toprule
Parameter & Parameter type & Description \\
\midrule
model\_id & INT & It is a unique identifier for a specific  ads ranking model. \\
model\_type & STRING & It is a unique identifier for a specific ads ranking model type. \\
enable\_flag & BOOL & It determines whether or not the cache is enabled. \\
cache\_ttl & INT & It is used that specifies the duration for which embeddings are valid in the cache. \\
\bottomrule
\end{tabular}
\caption{ERCache Configuration Parameters}
\label{tab:ercache_params}
\end{table*}

\begin{figure*}[htbp]
  \centering
  \includegraphics[width=.9\textwidth]{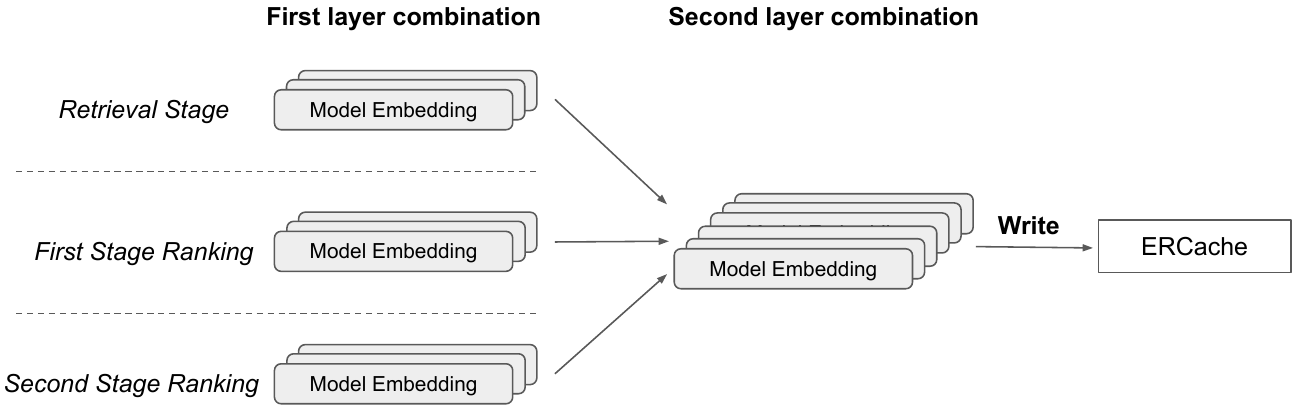}
  \caption{Combination technique of \sysname{}.}
  \label{fig:group}
\end{figure*}

\subsection{Customized cache configurations}
We chose the TTL-based eviction policy due to its alignment with user access patterns and time-based prioritization, which is more suitable for this cache design compared to LRU or other policy-driven approaches. 

This approach ensures that items are evicted from the cache based on their age, which aligns with the natural decay of user interest in content over time. Additionally, it allows us to prioritize items based on their recency, ensuring that the most recently accessed items are kept in the cache for longer periods of time. This approach also simplifies the cache management process, as it eliminates the need for complex heuristics or algorithms to determine which items to evict. 

Overall, the TTL-based eviction policy provides a simple and effective solution for managing the cache and ensuring that it remains relevant and useful to users.

ERCache offers caching capabilities for 
individual model IDs or model types. 
Customers have the flexibility to 
enable caching based on their specific needs. 
The ERCache configuration includes various parameters, as shown in Table~\ref{tab:ercache_params}.

\subsection{Update combination}
ERCache employs a two-layer update combination mechanism 
to minimize the number of cache write requests per user 
across multiple ranking stages, shown in Figure~\ref{fig:group}. 
By consolidating user embeddings from various ranking models
across multiple ranking stages
into a single request, 
rather than having one request per model embeddings per stage, 
we significantly reduce the write QPS on the \sysname. 

\subsection{Asynchronous write}
After grouping all cache write requests into one single request,
we send the write request to ~\sysname{} asynchronously.
The asynchronous operation moves write
out ot the critical path and
does not impact the e2e latency.

\subsection{Regional consistency}
\sysname{} guarantees the regional consistency
through its internal memcache system.
Since most requests are routed to the same region
as their previous serving for good locality, 
both the request and cache remain in the same region most of the time, 
ensuring efficient data access 
and minimizing latency.

\subsection{Reliability}
\sysname{} may face cascading effects due to 
traffic oscillations, regional outages, and site events, 
leading to increased load and reduced performance. 
To enhance system reliability, a rate limiter has been implemented. 
This rate limiter filters requests based 
on regional thresholds if there is a sudden spike in QPS.


\section{Evaluation}
\begin{table*}[htbp]
  \begin{tabular}{cccccc}
    \toprule
    Predictor task & Ranking stage & Direct cache TTL &
    Computation resource savings & E2E p99 latency diff \\
    \midrule
    CVR&First & 5 minutes & 44\% & -0.4\%\\
    CVR&First & 5 minute & 51\% & -0.11\%\\
    CTR&First & 5 minutes & 43\% & -0.04\%\\
    CTR&Second & 5 minutes & 64\% & -0.03\%\\
    CVR&Second & 1 minutes & 52\% & -0.4\%\\
    \bottomrule
  \end{tabular}
  \caption{The \sysname{} (direct cache) performance on ads ranking models at Meta.}
  \label{tab:direct}
\end{table*}

\begin{table*}[htbp]
  \begin{tabular}{ccccc}
    \toprule
    Predictor task & Ranking stage & Failover cache TTL & Fallback rate w/o cache & Fallback rate w/ cache\\
    \midrule
    CVR&Retrival & 1 hour & 0.7\% & 0.3\%\\
    CTR&Retrival & 1 hour & 0.6\% & 0.1\%\\
    CVR&First & 1 hour & 5.9\% & 0.1\%\\
    CVR&First & 1 hour & 6.5\% & 0.1\%\\
    CTR&First & 1 hour & 1.5\% & 0.5\%\\
    CTR&First & 1 hour & 1.4\% & 0.1\%\\
    CTR&Second & 2 hours & 0.05\% & 0.01\%\\
    CVR&Second & 2 hours & 0.1\% & 0.04\%\\
    \bottomrule
  \end{tabular}
  \caption{The \sysname{} (failover cache) performance on ads ranking models at Meta.}
  \label{tab:failover}
\end{table*}

In this section, we evaluate \sysname to answer the following questions:
\begin{enumerate}
    \item [1.] How much computational resources can \sysname{} save?
    \item [2.] What's the impact of \sysname{} on service SLAs?
    \item [3.] How can \sysname{} affect model performance with 
    different cache TTL?
    \item [4.] What is the effect of varying cache TTL settings on cache performance?
    \item [5.] What are the serving costs of \sysname?
    \item [6.] Is \sysname{} reliable when faced with cascading effects, such as sudden changes in traffic or system failures?
\end{enumerate}


\subsection{Experimental setup}
In this study, all experiments were conducted on industrial datasets 
using A/B testing in our production system. 

To measure the effcetiveness of \sysname, 
we compare the computational resources and 
key service SLAs for 
enabling and disabling ads ranking models at Meta.

Specifically, we measure computational resources 
by the power consumed during model inference 
using both CPU and GPU. 
In terms of service SLAs, 
we focus on the key metrics that 
impact our system's performance, 
including end-to-end (e2e) p99 latency
and model fallback rate.
Additionally, we evaluate model performance based on 
Normalized Cross Entropy (NE).

To further understand the performance of \sysname, 
we measure the cache hit rate with 
different cache TTL settings. 
We also evaluate the serving cost of \sysname{} 
and its reliability during cascading effects.

\subsection{Computational resource evaluation}
As mentioned earlier, we measure the power usage 
for a model and compare the change with and without 
direct cache to evaluate the effectiveness of \sysname{} 
in reducing computational resources.
From Table~\ref{tab:direct},
we find \sysname{} can significantly reduce computational resource 
usage by 42\% to 64\%, 
depending on the cache TTL settings.
Furthermore, we note that 
the power savings achieved by \sysname{} 
vary across different models, 
due to their distinct access patterns and model profiles.

\subsection{Service SLAs evaluation}
To understand the imapct of \sysname{} on service SLAs,
we evaluate e2e p99 latency and model fallback rate
with \sysname{} enabled.

\textbf{E2E p99 latency.}
According to Table~\ref{tab:direct}, 
we achieved an average reduction of 0.2\% 
in end-to-end p99 latency.
This improvement is attributed to 
the decrease in the number of model inference requests 
and reduced workload in the ads recommendation systems.
Notably, we did not observe 
any NE loss for the models with direct cache enabled, 
using the cache TTL shown in Table~\ref{tab:direct}.
Notably, we did not observe any NE loss 
for the models with direct cache enabled, 
using the cache TTL shown in Table~\ref{tab:direct}.

\textbf{Model fallback rate.}
Table~\ref{tab:failover} shows that 
the failover cache effectively reduces 
the fallback rate for ads ranking models, 
with an average reduction of 79.6\%. 
The most notable improvement is observed 
in the CVR model at the first ranking stage, 
where the fallback rate decreased from 6.5\% to 0.1\%.

\subsection{Impact of cache TTL}
To further understand the impact of cache TTL,
we evaluate model performance and cache performance
using different cache TTL settings.

\textbf{Model performance evaluation.}
\begin{table}[htbp]
  \begin{tabular}{cc}
    \toprule
    Direct cache TTL & NE difference \\
    \midrule
    30 seconds & 0.002\% \\
    1 minute & -0.001\% \\
    2 minutes & -0.007\% \\
    5 minutes & 0.003\% \\
    10 minutes & 0.06\% \\
    \bottomrule
  \end{tabular}
  \caption{The impact of cache TTL on model performance.}
  \label{tab:model}
\end{table}
We investigate the relationship between model performance (NE) 
and cache TTL by comparing the NE difference 
between enabling and disabling direct cache with 
varying cache TTLs, ranging from 30 seconds to 10 minutes. 
We find that the model's performance starts 
to degrade when the cache TTL is set to 10 minutes or higher, 
as shown in Table~\ref{tab:model}. 
It is important to note that a lower NE value 
indicates better model performance.

\textbf{Impact of cache TTL on cache performance.}

\begin{figure}[tbp]
  \centering
  \includegraphics[width=.45\textwidth]{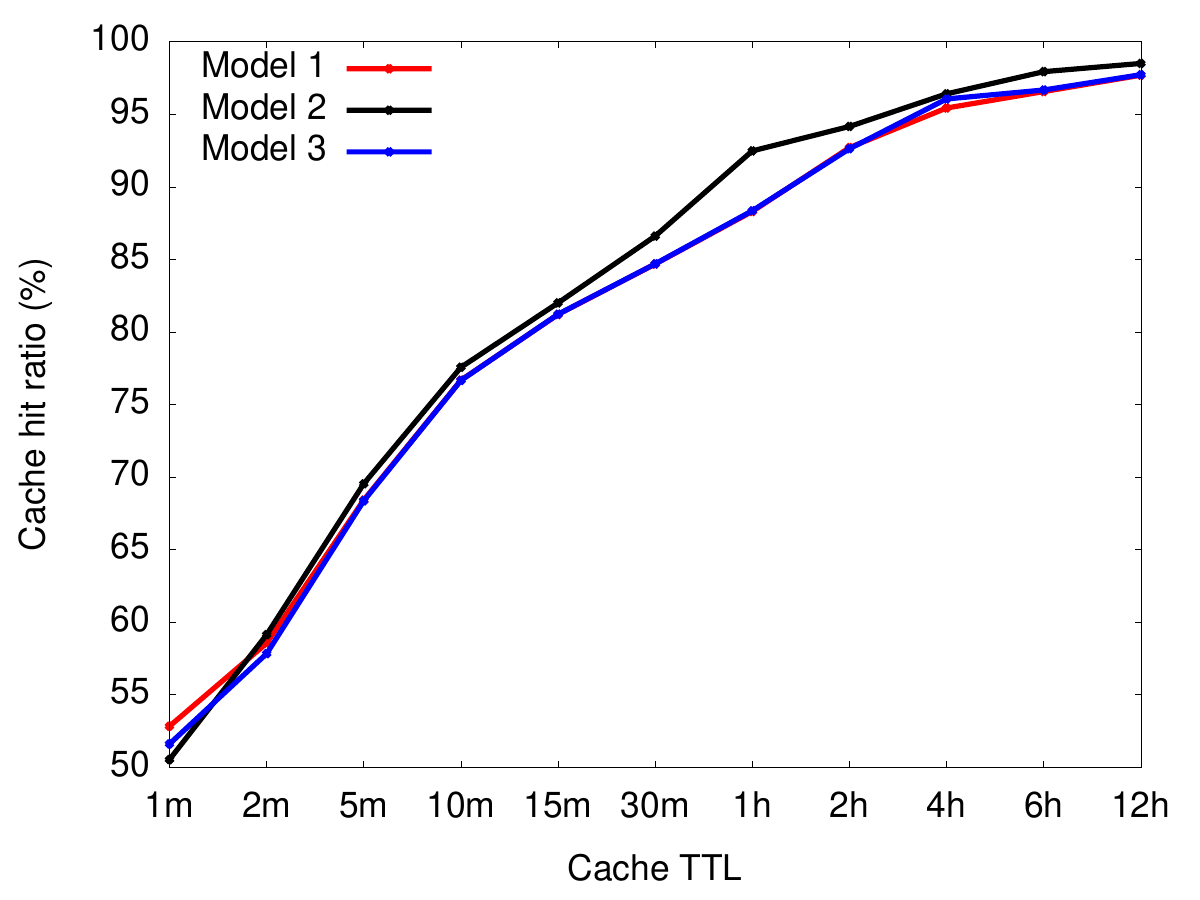}
  \caption{Impact of cache TTL on direct cache.}
  \label{fig:cache-hit}
\end{figure}

Figure ~\ref{fig:cache-hit} shows the impact of cache TTL over direct cache.
On average, a 1-minute TTL yields a 51.6\% cache hit rate, 
while a 5-minute TTL results in a 68.7\% cache hit rate. 
A 1-hour TTL achieves an impressive 89.7\% cache hit rate, 
and extending the TTL to 6 hours leads to a remarkable 97.1\% cache hit rate.
Furthermore, a 12-hour TTL can achieve an outstanding 97.9\% cache hit rate.

\textbf{Optimizing cache TTL settings in production.}
In practice, we typically set a shorter TTL for the direct cache
and a longer TTL for the failover cache. 
This is because we prioritize maintaining model performance 
by using a short TTL in the direct cache, 
while the failover cache is designed to compensate 
for failed model inference requests 
and is less concerned with data freshness, 
so a longer TTL can be used.

~\subsection{Serving cost of \sysname}
We evaluate the serving cost of \sysname{} from
QPS, latency, and bandwidth.

\textbf{QPS}.
\begin{figure}[tbp]
  \centering
  \includegraphics[width=.45\textwidth]{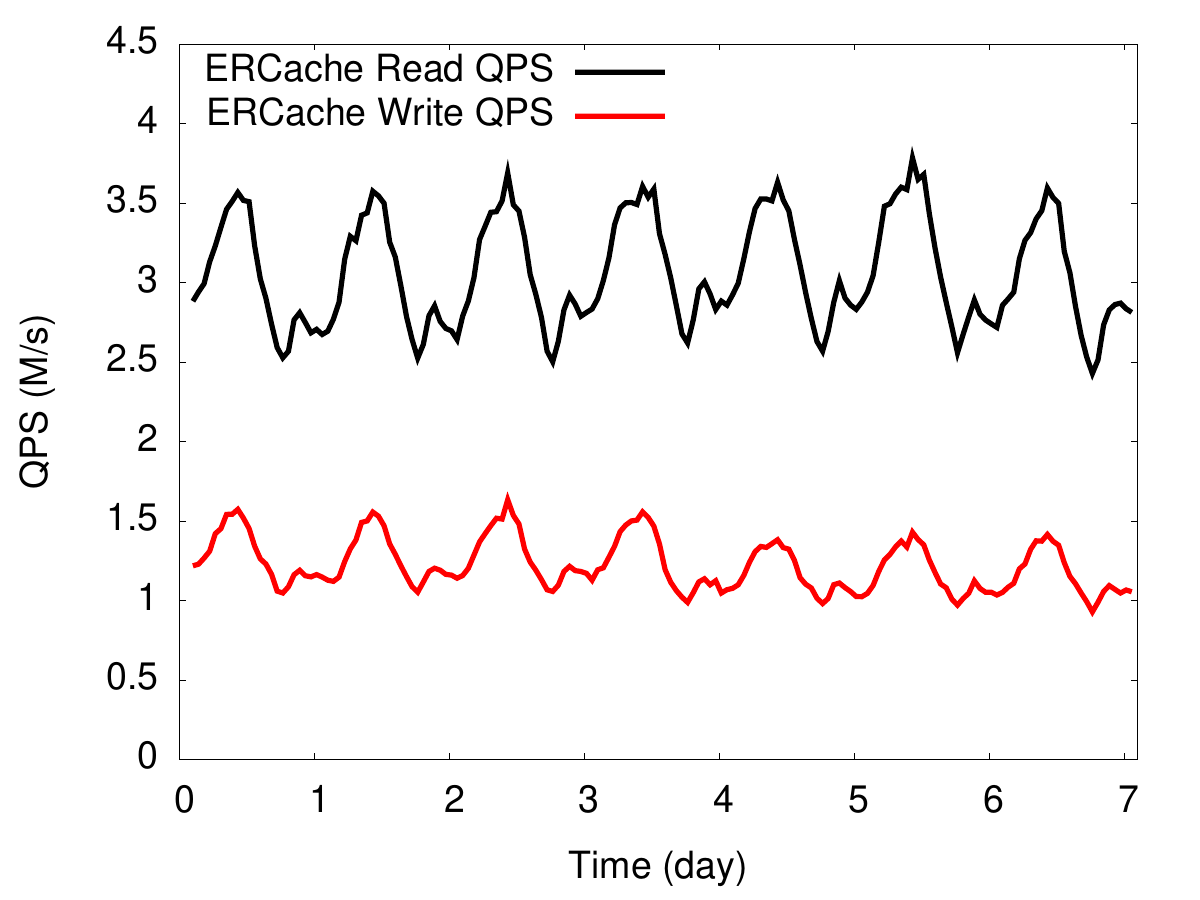}
  \caption{\sysname{} QPS over a week period.}
  \label{fig:qps}
\end{figure}
Figure~\ref{fig:qps} shows the write QPS, 
which ranges from 0.93 M/s to 1.63 M/s, 
and the read QPS, which varies between 2.43 M/s and 3.778 M/s. 
By applying the caching grouping technique, 
we have successfully reduced the number of cache reads and writes. 
If we were to support 30 models without this grouping technique, 
the QPS would increase by at least 30x.

\textbf{Latency}.
\begin{figure}[tbp]
  \centering
  \includegraphics[width=.45\textwidth]{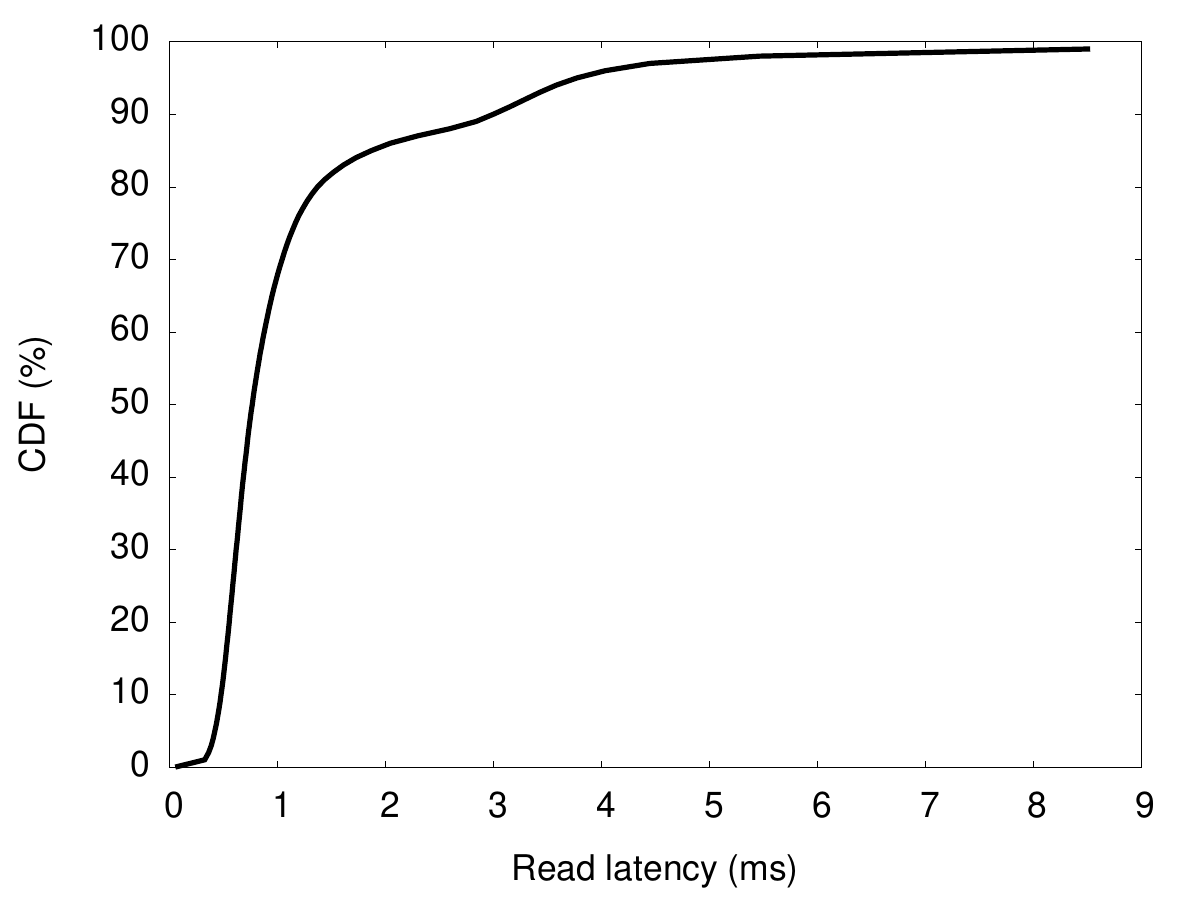}
  \caption{CDF of \sysname{} read latency.}
  \label{fig:latency}
\end{figure}
Figure~\ref{fig:latency} displays the CDF of read latency in \sysname{}. 
The results show that 50\% of read requests 
have a latency of less than 1 ms, 
while 80\% have a latency of less than 2 ms. 
The p50 read latency is 0.77 ms and 
the p99 read latency is 8.47 ms. 
Most read requests can be completed within 10 ms.
The low latency of \sysname{} 
does not hurt the performance of ads recommendation systems.

Since we use asynchronous write to \sysname{}, 
we do not prioritize the write latency. 

\textbf{Write bandwidth}.
\begin{figure}[tbp]
  \centering
  \includegraphics[width=.45\textwidth]{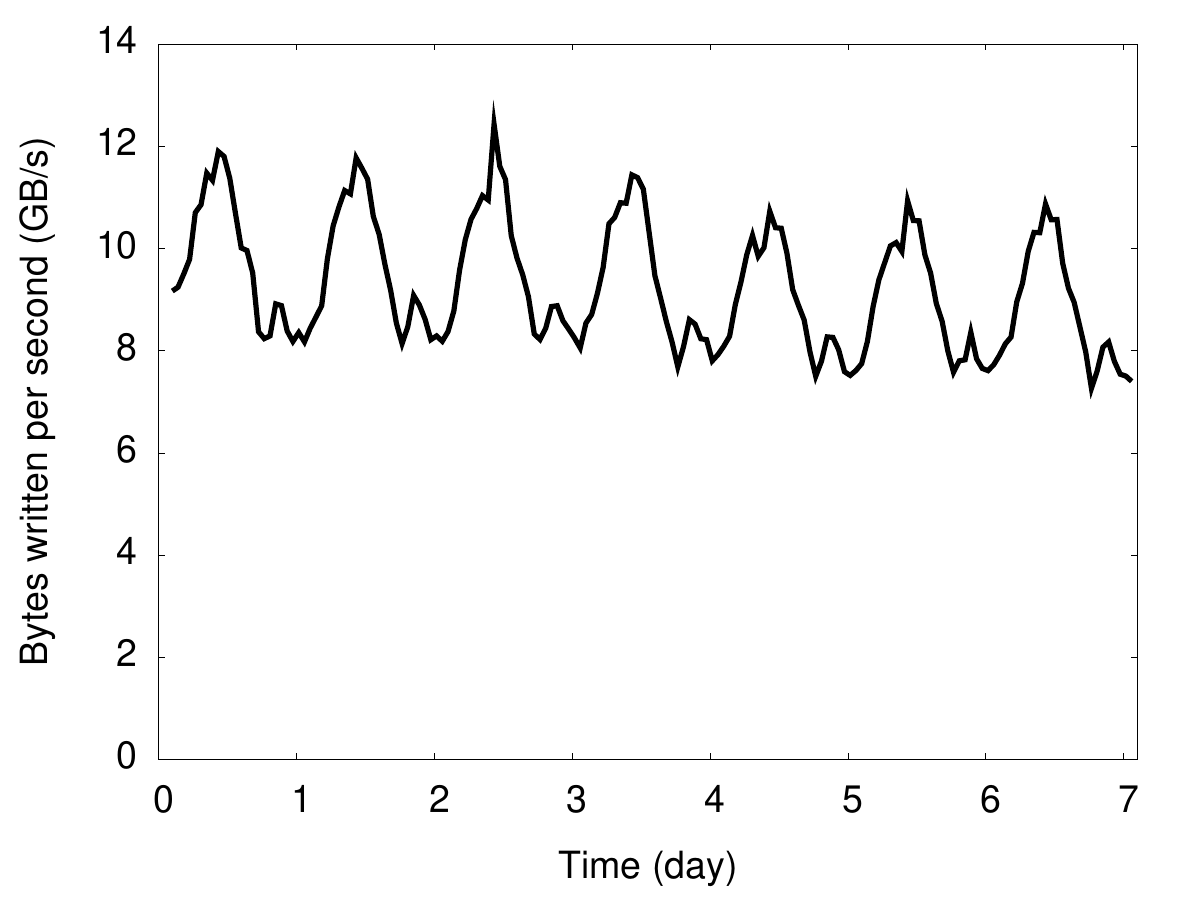}
  \caption{Bytes bandwidth of \sysname{} over a week period.}
  \label{fig:memsize}
\end{figure}
The write bandwidth of \sysname{} varies between
7.26 GB/s and 12.43 GB/s, with an average of 9.16 GB/s, shown in Figure~\ref{fig:memsize}.
We do not discuss the read throughput 
as it is relatively inexpensive in memory.

~\subsection{Reliability of ~\sysname{}}
To assess the reliability of ~\sysname{}, 
we conducted a drain test on one region and 
monitored its performance during this special situation. 
The drain test involved intentionally taking down a data center/region 
to simulate a disaster scenario, 
such as a fire or power outage.

\begin{figure}[tbp]
  \centering
  \includegraphics[width=.45\textwidth]{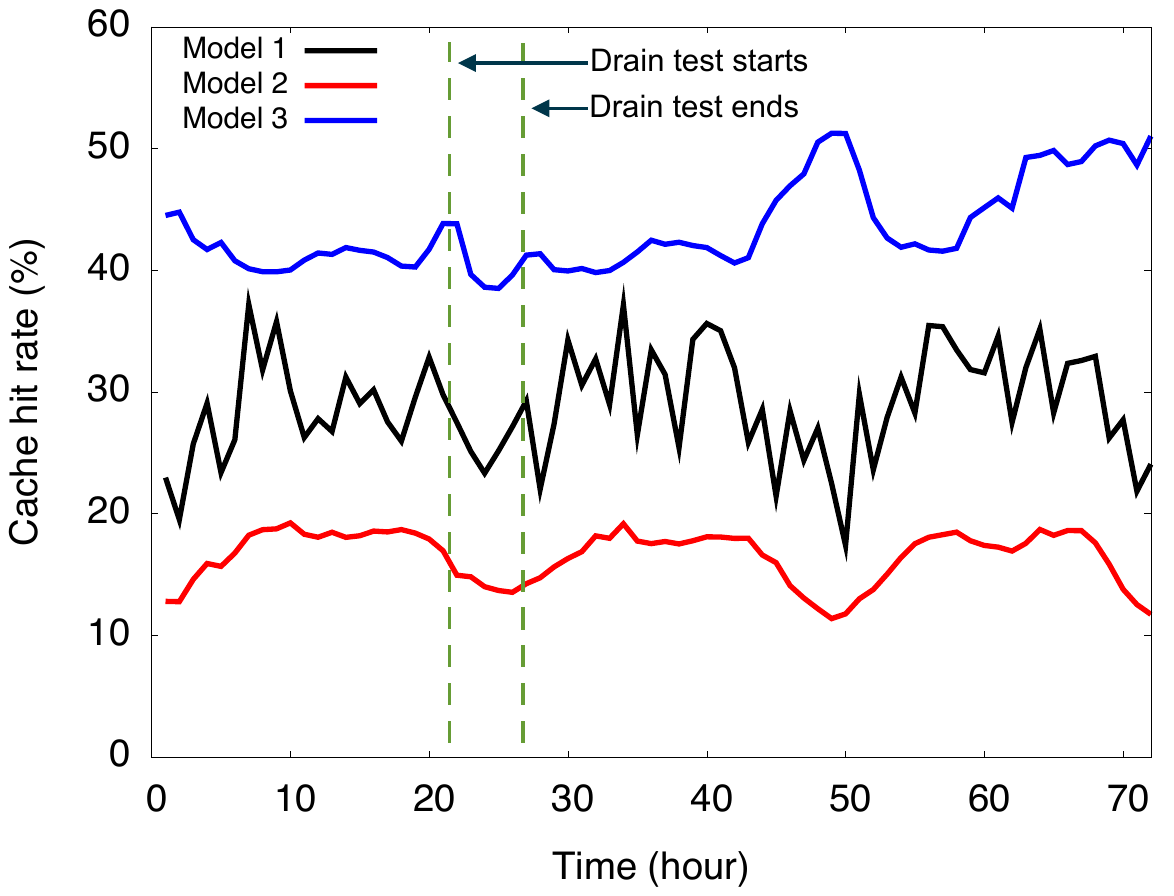}
  \caption{6-hour drain test for \sysname{}}
  \label{fig:drain}
\end{figure}

We ran a 6-hour drain test on one region out 
of 13 main regions. 
Figure\ref{fig:drain} presents the results 
of the reliability test. 
The drain test began at hour 21 and ended at hour 26. 
During the test period, we did not observe any unusual changes 
in \sysname's primary metrics. 
The cache hit rate of ~\sysname{} remained stable throughout the period.
The results demonstrate that ~\sysname{} 
can withstand severe situations, 
such as cascading effects, and exhibits good reliability.

~\subsection{Key takeaways}
\sysname{} has been deployed at Meta for over two years, 
providing support to more than 30 ranking models 
and ensuring improved model performance 
in accordance with service SLAs.

The success of \sysname{} demonstrates that 
\begin{enumerate}
    \item [1.] Model inference is not necessary for every ads request, 
despite the importance of embedding freshness to model performance. 
    \item [2.] The triangular relationship between model complexity, embedding freshness, 
and service SLAs is useful and reasonable.
It can serve as a reference for other researchers and engineers developing models and systems.
    \item [3.] The serving cost of \sysname{} is not expensive 
    due to its low QPS, latency, and bandwidth.
    \item [4.] \sysname{} can be easily applied to other ads recommendation systems in large-scale social networks or other areas with similar access patterns to Meta.
\end{enumerate}

\section{related work}
~\textbf{Training Optimization}
Recent publications focus on optimizing DRAM cache 
and GPU resident cache utilization for training purposes.
HierPS~\cite{zhao2020distributed} is a
distributed GPU hierarchical parameter server 
for massive scale deep learning ads systems
with 3-layer hierarchical storage 
including GPU HBM, CPU memory and SSD.
AIBox~\cite{zhao2019aibox} is
a centralized system to train CTR
models with tens-of-terabytes-gb
by SSDs and GPUs.
While they prioritize training optimization, 
our focus is on design of caching systems for efficient 
and reliable model inference.

~\textbf{Embedding optimization}
AdaEmbed~\cite{lai2023adaembed} is a complementary system, 
to reduce the size of embeddings needed for the same accuracy 
via in-training embedding pruning.
It prioritizes embeddings with high runtime access 
frequencies and large training gradients, 
dynamically pruning less important ones to optimize per-feature embeddings.
AdaEmbed targets embedding optimization during the training phase, 
which is a distinct area of caching research within our work.

~\textbf{Inference optimization}
Fleche~\cite{xie2022fleche} presents a comprehensive cache 
scheme with detailed designs 
for efficient GPU-resident embedding caching.
UGACHE~\cite{song2023ugache} introduces 
a novel factored extraction mechanism that 
mitigates bandwidth congestion to 
fully utilize high-speed cross-GPU interconnects.
RECom~\cite{pan2023recom} 
proposes the first ML compiler designed to optimize 
the massive embedding columns in recommendation models on the GPU.
EVStore~\cite{kurniawan2023evstore} is a 3-layer table lookup
system using both structural regularity 
in inference operations and 
domain-specific approximations 
to provide optimized caching.
However, these works focus on optimizing the performance of model inference, 
whereas \sysname{} targets the caching system 
before sending requests to model inference.

~\textbf{Cache case study}
Twitter has published an analysis of its internal 
caching system~\cite{yang2021large}. 
The paper aims to characterize cache workloads 
based on traffic patterns, TTL, 
popularity distribution, and size distribution. 
However, this analysis is too broad and not specifically 
tailored to ads recommendation systems. 
As a result, ads recommendation systems may not 
find much value in such a general analysis.

\section{Conclusion}
We introduce \sysname, 
a caching framework specifically 
designed to efficiently and reliably 
manage large-scale user representations. 
ERCache helps alleviate computational resource limitations 
for increasingly complex models while
ensuring that onboarding complex models 
meets SLAs.

By utilizing a direct and failover cache system 
alongside customized eviction policies, 
~\sysname{} effectively balances model complexity, 
embedding freshness, and SLAs, 
despite the inherent staleness introduced by caching.

~\sysname{} has been successfully deployed 
in Meta’s production systems for over half a year, 
supporting more than 30 ad ranking models. 
This deployment has significantly reduced 
computational resource requirements 
while maintaining service SLAs.

Apart from the practical contributions, 
the triangular relationship identified in this study, 
along with the success of \sysname{}, 
provides a valuable reference for the 
design and research of ad recommendation systems.

\bibliographystyle{ACM-Reference-Format}
\bibliography{sample-base}


\end{document}